\documentclass{aa}
\usepackage[latin1]{inputenc}
\usepackage[english]{babel}
\usepackage{array}
\usepackage{bm}
\usepackage{units}
\usepackage{amsmath}
\usepackage{amssymb}
\usepackage{graphicx}
\usepackage[usenames]{color}
\usepackage{psfrag}
\usepackage{float} 
\usepackage{graphicx}
\usepackage{epsfig}
\usepackage{natbib}

\begin{document}

\title{Energy production in varying $\alpha$ theories}

\author{Lucila Kraiselburd$^{1}$ \thanks{fellow of CONICET}  \and 
        Marcelo Miller Bertolami$^{2,3}$ \thanks{member of  the Carrera del Investigador
    Cient\'{\i}fico y Tecnol\'ogico, CONICET} \and
        Pablo Sisterna$^{4}$ \and  H\'{e}ctor Vucetich$^{1}$ } 

\institute{Grupo de Gravitaci\'{o}n, Astrof\'{i}sica y Cosmolog\'{i}a, Facultad de Ciencias Astron\'{o}micas y Geof\'{\i}sicas,
  Universidad Nacional de La Plata, Paseo del Bosque S/N, cp 1900 La
  Plata, Argentina\\ 
\and 
Instituto de Astrof{\'\i}sica La Plata \\
\and
Grupo de Evoluci\'{o}n estelar y pulsaciones, Facultad de Ciencias Astron\'{o}micas y 
Geof\'{\i}sicas, Universidad Nacional de La Plata, Paseo del Bosque S/N, cp 1900 La
  Plata, Argentina\\  
\and
Facultad de Ciencias Exactas y Naturales, Universidad Nacional de Mar del Plata, Funes 3350,
    cp 7600 Mar del Plata, Argentina \\ 
\email{lkrai, mmiller, vucetich@fcaglp.fcaglp.unlp.edu.ar,
  sisterna@mdp.edu.ar}} 

 
\abstract {} {On the basis the theoretical model proposed by Bekenstein for $\alpha$'s variation, we analyze the equations that describe the energy exchange between matter and both the electromagnetic and the scalar fields.}  {We determine how the energy flow of the material is modified by the presence of a scalar field. We estimate the total magnetic energy of matter from the ``sum rules techniques''. We compare the results with data obtained from the thermal evolution of the Earth and other planets.}  { We obtain stringent upper limits to the variations in $\alpha$ that are comparable  with those obtained from atomic clock frequency variations.}  {Our constraints imply that the fundamental length scale of Bekenstein's theory ``$\ell_B$'' cannot be larger than Planck's length ``$\ell_P$''.}

\authorrunning{Kraiselburd et al.}

\titlerunning{Variation in $\alpha$ and energy production}

\maketitle

  \section{Introduction}
  \label{sec:Intro}

The time variation in the fine structure constant has been studied several times since first being proposed by \citet{gamow67}. Observational 
upper bounds on its time variation as well as several theoretical frameworks that consider $\alpha$ as a dynamical field have been published (an exhaustive list can be found in \citep{landau02:tesis,uzan03} and references there in). Although still disputed, the claim that $\alpha$  was smaller in the past is an exciting perspective, \citep{MurphyWebbFlambaum2003}. 

Beckenstein's theory \citep{Beckens82}, which is based on a number of minimal hypothesis of highly accepted physical principles, is in a sense representative of many low energy theories inspired by grand unification schemes.
In this work, we derive equations that govern the energy exchange between matter, the scalar field, and the electromagnetic field. Although we do not analyze the precise mechanism of energy release, we assume that the work done by the scalar field is radiated away in an efficient way, as for the rotochemical heating of neutron stars due to the spin down of the star \citep{Reisenegger1995,FernandezReisenegger2005}.  

In section \ref{sec:alfadot}, we briefly review Beckenstein's theory, as well as the cosmological time evolution of $\alpha$ that it predicts. In section \ref{sec:Edot}, we derive a generalized version of the Poynting theorem for the electromagnetic field, and from the conservation of the total energy-momentum tensor we find how the energy flow of matter is modified by the scalar field. In section \ref{sec:Eem-Mat}, we discuss the magnetic energy of matter using a simple nuclear model. In section \ref{sec:CoolEarth}, we study the thermal history of the Earth in the presence of Bekenstein's scalar field. We also describe in section \ref{sec:outerplanets} the results we obtained for the outer planets. Finally in section \ref{sec:conclusions} we summarize our conclusions.

\section{Time variation of $\alpha$ in Bekenstein's formalism}
\label{sec:alfadot}

We briefly review Bekenstein's formalism and its prediction for the cosmological time variation of $\alpha$. Although we consider galactic as well as terrestrial phenomena, we can nevertheless confidently assume that they track the cosmological evolution of $\alpha$, \citep{ShawBarrow2006}.   

\citet{Beckens82} proposes to modify Maxwell's theory by introducing  a field $\epsilon$ that dynamically describes the variation of $\alpha$. The
foundational hypothesis are the following \citep{Beckens82,landau02:tesis}:
\begin{enumerate}
\item \label{item:Maxwell} The theory must reduce to Maxwell's when
$\alpha=\mathrm{Cte}$.
\item The changes in $\alpha$ are dynamical (i.e. generated by a dynamical field $\epsilon$). 
\item The dynamics of the electromagnetic field, as well as the $\epsilon$ field can be obtained from a variational principle.
\item The theory must be locally gauge invariant.
\item The theory must preserve causality.
\item The action must be time reversal invariant.
\item \label{item:Planck} Planck's scale $\ell_P$ is the smallest length available in the theory.
\item \label{item:Einstein} Einstein's equations describe gravitation.
\end{enumerate}
  
String theories and the like in which there are other fundamental length scales, force us to set aside condition \ref{item:Planck}. These hypothesis uniquely lead to the action
\begin{equation}
 S = S_{\rm em} + S_\epsilon + S_m + S_G ,\label{equ:S-Bek} 
\end{equation}
where
\begin{eqnarray}
 S_{\rm em} &=& - \frac{1}{16\pi} \int F^{\mu\nu}F_{\mu\nu}
 \sqrt{-g} d^4x, \label{equ:Mod-S-em}\\
 S_\epsilon &=& -\frac{\hbar c}{2\ell_B} \int
 \frac{\epsilon^{,\mu}\epsilon_{,\mu}}{\epsilon^2} \sqrt{-g}
 d^4x, \label{equ:S-eps}
\end{eqnarray}
$ S_m $ and $S_G $ are the matter and gravitational field actions, respectively, $\ell_B$ is the so-called Bekenstein's fundamental length, and the metric here is $(-1,1,1,1)$. 

The main difference between Maxwell's and Bekenstein's theories is the connection between the vector potential and the electromagnetic field
\begin{equation}
  F_{\mu\nu} = \frac{1}{\epsilon} \left[(\epsilon A_\nu)_{,\mu} -
  (\epsilon A_\mu)_{,\nu}\right] \label{equ:Fmn-A}
\end{equation}
and the second kind of local gauge invariance implies that
\begin{eqnarray}
 \epsilon A'_\mu &=& \epsilon A_\mu + \chi_{,\mu}, \label{equ:TG-Bek}\\
 \nabla_\mu &=& \partial_\mu - e_0\epsilon A_\mu, \label{equ:DC-Bek}
\end{eqnarray}
as the gauge transformation and covariant derivative of the theory, respectively. The last equation defines the local value of the elementary electric charge (coupling constant)
\begin{equation}
 e(\bm{r},t) = e_0 \epsilon(\bm{r},t), \label{equ:e(r,t)}
\end{equation}
that is
\begin{equation}
 \epsilon = \left(\frac{\alpha}{\alpha_0}\right)^{\frac{1}{2}}.
\label{equ:eps-alfa} 
  \end{equation}
In what follows, we neglect the small spatial variations in $\alpha$ and focus on the cosmological variation, as we are interested in any secular energy injection of the scalar field on a planet such as the Earth. In our approximate analysis it is also enough to work in a flat space-time.

The field equations for the electromagnetic field and for $\epsilon$ are

\begin{subequations}\label{ecs:MovBeck}
 \begin{gather}
    \left(\frac{1}{\epsilon}F^{\mu\nu}\right)_{,\nu} = 4\pi j^\mu,
           \label{equ:Mov:em} \\
    \begin{split}
    \Box \ln\epsilon &= \frac{\ell_B^2}{\hbar c}
     \left[\epsilon\frac{\partial\sigma}{\partial\epsilon} - 
      \epsilon j^\mu A_\mu + \frac{1}{4\pi}\left(A_\mu
      F^{\mu\nu}\right)_{,\nu}\right]\\ 
	  &= \frac{\ell_B^2}{\hbar c}
    \left(\epsilon\frac{\partial\sigma}{\partial\epsilon} -
    \frac{F^{\mu\nu}F_{\mu\nu}}{8\pi}\right), \label{equ:Mov:epsilon}
    \end{split}
 \end{gather}
\end{subequations}
where $j^{\mu}=\sum(e_0/c\gamma)u^{\mu}(-g)^{-1/2}\delta^3[x^i-x^i(\tau)]$, $u_{i}^{\mu}$ is an ``standard estimate'' of the 4th velocity of each particle according to the model, and $\sigma$ is the energy density of matter, \cite{Beckens82}. $\Box$ is the covariant flat d'Alambertian
\begin{equation}
  \Box\phi = {\phi^{,\mu}}_{,\mu} = \eta^{\mu\nu} \phi_{,\mu,\nu}.\label{def:Box} 
\end{equation}
A note regarding the matter Lagrangian is in order: in \citep{Beckens82,Bekenstein:2002wz}, Bekenstein represents matter as an ensemble of classical particles. However, wherever quantum phenomena become important, as in white dwarfs or condensed matter physics, this is not a realistic description. It is also an inaccurate description at high energies (or on small length scales) because fermions have a ``natural length scale'', the particle Compton wave-length $\lambda_C = \hbar/mc$, which makes quite unrealistic any classical model at higher energies. In particular, several conclusions of \citet{Bekenstein:2002wz} have to be reconsidered.

In \citet{Beckens82}, it is shown that the cosmological equation of motion for $\epsilon$ is
\begin{equation}
  \frac{d\ }{dt} \left(a^3\frac{\dot{\epsilon}}{\epsilon}\right) = 
  -a^3 \frac{\ell_B^2}{\hbar c} \left[\epsilon
  \frac{\partial\sigma}{\partial\epsilon} -
  \frac{1}{4\pi}\left(\mathbf E^2 - \mathbf B^2\right)\right]. \label{equ:MovCosmoEps}
\end{equation}
In the non-relativistic regime, $\mathbf E^2 \gg \mathbf B^2$ and $\sigma \propto  \epsilon^2$, hence
\begin{equation}
  \frac{d\ }{dt} \left(a^3\frac{\dot{\epsilon}}{\epsilon}\right) =
  -a^3 \zeta_c \frac{\ell_B^2}{\hbar c} \rho_m c^2,
  \label{equ:MovCosmoEps:zeta-c} 
\end{equation}
where $\rho_m$ is the total rest-mass density of electromagnetically interacting matter and $\zeta_c$ is a parameter describing its ``electromagnetic content'', which is essentially the fractional contribution of the electromagnetic energy to the rest mass. A first estimation according to \cite{Bekenstein:2002wz} is

\begin{equation}
  \zeta_c \sim 1.2\times10^{-3} .\label{equ:zeta-c:num}
\end{equation}
Following the standard cosmological model, we assume that dark matter is to be electromagnetically neutral.

Given that $\rho_m \propto a^{-3}$, we can integrate Eq.(\ref{equ:MovCosmoEps:zeta-c}) obtaining 
\begin{displaymath}
 \frac{\dot{\epsilon}}{\epsilon} = - \zeta_c
 \left(\frac{\ell_B^2c^3}{\hbar}\right) \rho_m (t-t_c), 
\end{displaymath}
which can be written, using the usual cosmological notation, as follows
\begin{equation}
  \frac{\dot{\epsilon}}{\epsilon} = -\frac{3\zeta_c}{8\pi}
  \left(\frac{\ell_B}{\ell_P}\right)^2 H_0^2 \Omega_B
  \left[\frac{a_0}{a(t)}\right]^3 (t-t_c). \label{equ:Int1ra}
\end{equation}
The primordial nucleosynthesis standard model tells us that the integration constant $t_c$ must be very small in order not to spoil the agreement between theory and observation. Using WMAP values, we obtain the prediction for $(\dot{\alpha}/{\alpha})_0$ of
\begin{equation}
  \left(\frac{\dot{\alpha}}{H_0\alpha}\right)_0 = 1.3\times10^{-5}
  \left(\frac{\ell_B}{\ell_P}\right)^2. \label{equ:Pred}
\end{equation}
Any measurement with a precision such as $\sigma(\dot{\alpha}/H_0\alpha) \sim 10^{-5}$ is difficult to achieve, so the comparison between theory and experiment is a difficult task.

The same arguments can be applied to many theories with varying $\alpha$, such as Kaluza-Klein \citep{landau02:tesis} or string spired theories such as Damour-Polyakov's \citep{Damour:1994ya,Damour:1994zq}. 

\section{Energy transfer in Bekenstein's formalism}
\label{sec:Edot}

We study how energy is injected and then released in varying $\alpha$ theories, in order to look for observable consequences in the emission of astrophysical as well as geophysical systems. The energy momentum tensor in
Bekenstein's theory is defined as usual to be
\begin{equation}
 T^{\mu\nu} = 2c\frac{\delta S}{\delta g_{\mu\nu}}. \label{def:T_munu}
\end{equation}
Using $c=1$, the electromagnetic contribution then has the same form as in Maxwell's theory
\begin{equation}
 T_{\mu\nu}^{\rm em} = \frac{1}{4\pi} 
 \left[F_{\mu\lambda} {F_\nu}^\lambda - \frac{g_{\mu\nu}}{4}
 F_{\lambda\sigma}F^{\lambda\sigma}\right], \label{equ:T_mn:em}
\end{equation}
the difference lying in the connection between the vector potential and the field given in Eq.(\ref{equ:Fmn-A}).

The energy-momentum tensor of the scalar field $\epsilon$ is
\begin{equation}
 T^{\mu\nu}_\epsilon = \frac{\hbar }{\ell_B^2}
 \left(\frac{\epsilon^{,\mu}\epsilon^{,\nu}}{\epsilon^2} -
 \frac{1}{2} g^{\mu\nu}
 \frac{\epsilon^{,\alpha}\epsilon_{,\alpha}}{\epsilon^2}\right).
 \label{def:Tmn:epsilon} 
\end{equation}

In what follows, we use the redefined field
\begin{equation}
  \psi = \ln\epsilon. \label{def:phi}
\end{equation}

As we consider local phenomena, we can work in a locally inertial coordinate system. We denote the ``field part of the energy-momentum tensor'' as the scalar plus electromagnetic energy momentum tensor
\begin{equation}
  T_{\rm f}^{\mu\nu} = T^{\mu\nu}_{\rm em} + T^{\mu\nu}_\epsilon .
  \label{def:Tfield} 
\end{equation}
In terms of $\psi$ and replacing $g^{\mu\nu}$ with $\eta^{\mu\nu}$, we obtain
\begin{equation}
 \begin{split}
   T_{\rm f}^{\mu\nu} =& \frac{1}{4\pi} 
   \left[F^{\mu\lambda} {F^\nu}_\lambda - \frac{1}{4}\eta^{\mu\nu}
   F_{\lambda\sigma}F^{\lambda\sigma}\right]\\
   &+ \frac{\hbar }{\ell_B^2} \left(\psi^{,\mu}\psi^{,\nu} -
   \frac{1}{2}\eta^{\mu\nu}\psi_{,\alpha}\psi^{,\alpha} \right).
 \end{split}\label{equ:Tfield}
\end{equation}

The divergence of $T_{\rm f}$ is
\begin{equation}
  \begin{split}
    {T^{\mu\nu}_{\rm f}}_{,\nu} =& \frac{1}{4\pi}
    \left[{F^{\mu\alpha}}_{,\nu} {F^\nu}_\alpha + {F^{\mu\alpha}}
    {{F^\nu}_\alpha}_{,\nu} - \frac{1}{2}\eta^{\mu\nu} F^{\alpha\beta}
    {F_{\alpha\beta,\nu}} \right]\\
      &+ \frac{\hbar }{\ell_B^2}
    \left({\psi^{,\mu}}_{,\nu}\psi^{,\nu} +
    \psi^{,\mu}{\psi^{,\nu}}_{,\nu} - \eta^{\mu\nu}
    \psi_{,\alpha,\nu}\psi^{,\alpha} \right).
  \end{split}\label{equ:div:Tf}
\end{equation}

The equations of motion Eqs.(\ref{ecs:MovBeck}) are
\begin{subequations}
 \begin{gather}
   {F^{\alpha\nu}}_{,\nu} = 4\pi e^\psi j^\alpha +
   \psi_{,\nu}{F^{\alpha\nu}},\\
   {\psi^{,\nu}}_{,\nu} = \Box\psi = \frac{\ell_B^2}{\hbar }
   \left(\frac{\partial\sigma}{\partial\psi} -
   \frac{F^{\mu\nu}F_{\mu\nu}}{8\pi}\right),
 \end{gather}
\end{subequations}

which can be used in Eq.(\ref{equ:div:Tf}) obtaining
\begin{equation}
  \begin{split}
    {T^{\mu\nu}_{\rm f}}_{,\nu} =& \frac{1}{4\pi}
    [{F^{\mu\alpha}}_{,\nu} {F^\nu}_\alpha - {F^\mu}_\alpha
    \left(4\pi e^\psi j^\alpha + \psi_{,\nu}{F^{\alpha\nu}}\right)\\ 
    &- \frac{1}{2}\eta^{\mu\nu} F^{\alpha\beta} {F_{\alpha\beta,\nu}}]\\ 
    &+ \frac{\hbar }{\ell_B^2}
    [{\psi^{,\mu}}_{,\nu}\psi^{,\nu} +
    \psi^{,\mu}\frac{\ell_B^2}{\hbar } 
    \left(\frac{\partial\sigma}{\partial\psi} -
    \frac{F^{\mu\nu}F_{\mu\nu}}{8\pi}\right)\\ 
    &- \eta^{\mu\nu}
    \psi_{,\alpha,\nu}\psi^{,\alpha}].
  \end{split}\label{equ:div:Tf:Subs}
\end{equation}

This expression can be simplified using the homogeneous Maxwell equations 
\begin{equation}
  F_{\alpha\beta,\gamma} = - F_{\beta\gamma,\alpha} -
  F_{\gamma\alpha,\beta},\label{Maxpro} 
\end{equation}
which cancels out the first bracket. The first and last term in the second bracket also cancel out, thus we obtain for Eq.(\ref{equ:div:Tf:Subs}) the expression

\begin{equation}
  \begin{split}
    {T^{\mu\nu}_{\rm f}}_{,\nu} =& - e ^\psi j^\alpha {F^\mu}_\alpha\\ 
    &+\psi_{,\nu}\left(\eta^{\mu\nu}\frac{\partial\sigma}{\partial\psi} + 
  T^{\mu\nu}_{\rm em} -\frac{1}{16\pi}\eta^{\mu\nu} F_{\alpha\beta}F^{\alpha\beta}\right). 
  \end{split}\label{equ:div:Tf:Simp} 
\end{equation}

We add to both sides of the equation the divergence of the energy momentum tensor of matter ${T^{\mu\nu}_{\rm m}}_{,\nu}$ in order to find the energy transfer (according to the hypothesis 8 in Sect. \ref{sec:alfadot}, we assume that Einstein's equations hold unmodified for the gravitational field and hence that the total energy momentum tensor is conserved)
\begin{equation}
  \begin{split}
  &{T^{\mu\nu}_{\rm f}}_{,\nu} + {T^{\mu\nu}_{\rm m}}_{,\nu} = 0\\ &=
  {T^{\mu\nu}_{\rm m}}_{,\nu}  - e^\psi j^\alpha
  {F^\mu}_\alpha\\ 
  &+\psi_{,\nu}\left(\eta^{\mu\nu}\frac{\partial\sigma}{\partial\psi} +
  T^{\mu\nu}_{\rm em}-\frac{1}{16\pi}\eta^{\mu\nu} F_{\alpha\beta}F^{\alpha\beta}\right). 
  \end{split}
\end{equation}
This equation explicitly shows the energy transfer from the field $\epsilon$ to matter
\begin{equation}
  {T^{\mu\nu}_{\rm m}}_{,\nu} = e^\psi j^\alpha {F^\mu}_\alpha -
  \psi_{,\nu}\left(\eta^{\mu\nu}\frac{\partial\sigma}{\partial\psi} + T^{\mu\nu}_{\rm em}-\frac{1}{16\pi}\eta^{\mu\nu} F_{\alpha\beta}F^{\alpha\beta} \right), \label{div:Tm}
\end{equation}
which is the source of any observable effect. From
\begin{equation}
  \psi_{,\nu} = \frac{\epsilon_{,\nu}}{\epsilon} = \frac{1}{2}
  \frac{\alpha_{,\nu}}{\alpha}, \label{equ:fi:alfa}
\end{equation}
we find that the ``machian'' contribution to energy transfer is given by
\begin{equation}
  {{T^{\mu\nu}_{\rm m}}_{,\nu}}^{(machian)} = \frac{1}{2}
  \frac{\alpha_{,\nu}}{\alpha}
  \left(\eta^{\mu\nu}\frac{\partial\sigma}{\partial\psi}
  +T^{\mu\nu}_{\rm em}-\frac{\eta^{\mu\nu}}{16\pi} F_{\alpha\beta}F^{\alpha\beta} \right).
  \label{div:Tm:Mach} 
\end{equation}
 Using Bekenstein's notation, that is, if the time-space components of $e^{\psi}F^{\mu\nu}$ are identified with $\mathbf E$ while space-space components are identified with $\mathbf B$, the contribution then takes the form 
\begin{equation}
  \begin{split} 
    {{T^{0\nu}_{\rm m}}_{,\nu}}^{(machian)} =& -\dot\psi\frac{\partial\sigma}{\partial\psi}+ e^{-2\psi}\mathbf\nabla\psi . \mathbf S+\dot\psi e^{-2\psi}\frac{(\mathbf B^2 + \mathbf E^2)}{8\pi}\\
    &+\frac{e^{-2\psi}\dot\psi}{8\pi} (\mathbf B^2 - \mathbf E^2)\\
    &= -\dot\psi\frac{\partial\sigma}{\partial\psi}+ e^{-2\psi}\mathbf\nabla\psi . \mathbf S+\dot\psi e^{-2\psi}\frac{\mathbf B^2}{4\pi},
  \end{split}\label{div:Tm:Mach0} 
\end{equation}
where $\mathbf S=\frac{\mathbf E \times \mathbf B}{4\pi}$. Then, the component $0$ of Eq.(\ref{div:Tm}) reads
\begin{equation}
{T^{0\nu}_{\rm m}}_{,\nu} = \mathbf j . \mathbf E - e^{-2\psi}\frac{\mathbf B^2\dot\psi}{4\pi}  - e^{-2\psi}\mathbf\nabla\psi . \mathbf S +\dot\psi\frac{\partial\sigma}{\partial\psi}. \label{div:Tm0}
\end{equation}

 An implicit assumption of our previous analysis and algebra is the generalized Poynting theorem. In its standard version, it involves only electromagnetic terms, while in our case it also involves the interaction between the electromagnetic and scalar fields given by
\begin{equation}
  \begin{split}
    {T^{\mu\nu}_{\rm em}}_{,\nu} =& \frac{1}{4\pi}[{F^{\mu\alpha}}_{,\nu} {F^\nu}_\alpha - {F^\mu}_\alpha\left(4\pi e^\psi j^\alpha + \psi_{,\nu}{F^{\alpha\nu}}\right)\\
    &- \frac{1}{2}\eta^{\mu\nu} F^{\alpha\beta} {F_{\alpha\beta,\nu}}]. 
  \end{split}\label{equ:poyntingref}
\end{equation}
Using again Eq.(\ref{Maxpro}),
\begin{equation}
  \begin{split}
  {T^{\mu\nu}_{em}}_{,\nu} =& -e ^\psi j^\alpha {F^\mu}_\alpha\\ 
  &+ \psi_{,\nu}\left[\frac{{F^\mu}_{\alpha}F^{\nu\alpha}}{4\pi}-\eta^{\mu\nu}\frac{F_{\alpha\beta}F^{\alpha\beta}}{16\pi}+\eta^{\mu\nu}\frac{F_{\alpha\beta}F^{\alpha\beta}}{16\pi}\right], 
  \end{split}\label{equ:div:Tf:Simp1poyn} 
\end{equation}
\begin{equation}
  {T^{\mu\nu}_{em}}_{,\nu} = - e ^\psi j^\alpha {F^\mu}_\alpha +\psi_{,\nu}(T^{\mu\nu}_{em}+\eta^{\mu\nu}\frac{F_{\alpha\beta}F^{\alpha\beta}}{16\pi}). 
\label{equ:div:Tf:Simp2poyn} 
\end{equation}
Then,
\begin{equation}
  \begin{split}
  {T^{0\rho}_{em}}_{,\rho} =& -\mathbf E \cdot \mathbf j +e^{-2\psi}\frac{(\mathbf E^2 + \mathbf B^2)}{8\pi}\dot\psi\\
  &+ e^{-2\psi}\mathbf S . \nabla\psi -\frac{e^{-2\psi}\dot\psi}{8\pi} (\mathbf B^2 - \mathbf E^2),
  \end{split}\label{equ:div:Tf:Simp3poyn} 
\end{equation}
\begin{equation}
  \begin{split}
    {{T_{em}}^{0\rho}}_{,\rho} =& \frac{\partial u_{em}}{\partial t} +  \mathbf\nabla .e^{-2\psi} (\frac{\mathbf E \times \mathbf B}{4\pi})\\ 
    &= -\mathbf E \cdot \mathbf j +\frac{e^{-2\psi}\mathbf E^2}{4\pi}\dot\psi+e^{-2\psi}\mathbf S . \nabla\psi,
  \end{split}\label{poynting}
\end{equation}
where ${{T}_{em}}^{00}{,_0}=  (\partial u_{em})/\partial t$, the electromagnetic energy is $u_{em}=e^{-2\psi}(\mathbf E^2 + \mathbf B^2)/(8\pi)$, ${{T_{em}}^{0i}}_{,i}= \mathbf\nabla .e^{-2\psi}(\frac{\mathbf E \times \mathbf B}{4\pi})= \mathbf\nabla .e^{-2\psi}\mathbf S$, and $\mathbf S$ is the Poynting vector. We note that this result is independent of the details of the gravitational and matter Lagrangians, as well as their interacting terms with the electomagnetic field. In particular, it holds independently of the details of the interaction of matter with the scalar field. We recall that the usual interpretation of the first term in the right hand side of Eq.(\ref{poynting}) is the work done by the electromagnetic field on matter. In the same fashion, we may interpret the second and last term as the work done by the electromagnetic field on the scalar field. An analog phenomenon would be that given by the work done by an increasing Newton constant $G$ on a planet augments its pressure and thus compresses it \citep{JofreReiseneggerFernandez2006}.
 
 We estimate the electrostatic contribution to the matter energy. In a non-relativistic system such as a light atom or nuclei, the electromagnetic energy is given by the electrostatic field that satisfies the equation
\begin{equation}
 \bm{\nabla\cdot}\bm{E}e^{-2\psi} = 4\pi\rho^0_{\rm em},\label{equ:Poisson-Bek} 
\end{equation}
where $\rho^0$ is the reference charge density. In the limit where $\alpha$ varies only cosmologically we have
\begin{equation}
  \bm{\nabla\cdot}{\bm E} = 4\pi e^{2\psi}\rho^0_{\rm em}, 
\end{equation}
whose solution is
\begin{equation}
  \bm{E} = e^{2\psi}\bm{E}_0,
\end{equation}
where $\bm{E}_0$ is the electrostatic reference field defined for $e^\psi=1$.
We can evaluate the electromagnetic energy density to be
\begin{equation}
  u_{em} = e^{-2\psi}\frac{(\mathbf B^2+\mathbf E^2)}{8\pi}=e^{2\psi} u_{em}^0,
\end{equation}
and the temporal variation
\begin{equation}
  \dot{u}_{em} = 2\dot{\psi}u_{em}+e^{2\psi}\dot{u}^{0}_{em} =\frac{\dot{\alpha}}{\alpha}u_{em}+e^{2\psi}\dot{u}^{0}_{em}. \label{equ:TV-E:alfa}
\end{equation}

If there were no scalar injection of energy and $\dot{u}^{0}_{em}\approx 0$, the Poynting theorem Eq.(\ref{poynting}) and the energy variation given by Eq.(\ref{equ:TV-E:alfa}) would lead to

\begin{equation}
 2\dot{\psi}u_{em}=2\dot{\psi}e^{-2\psi}\frac{(\mathbf B^2+\mathbf E^2)}{8\pi}=  -\bm{j\cdot E}+\dot{\psi}e^{-2\psi}\frac{\mathbf E^2}{4\pi} \label{equ:Point2}
\end{equation}
or
\begin{equation}
\bm{j\cdot E} = -\frac{\mathbf B^2}{4\pi} \dot{\psi}e^{-2\psi}. \label{equ:Ohm-Vac} 
\end{equation}

As we consider phenomena where the motion of matter is negligible, taking the first index as $0$ is equivalent to projecting along the fluid four-velocity. In addition, the total time derivative $d/dt=\partial/\partial t + \mathbf v . \mathbf\nabla$ will be equal to the partial time derivative $\partial/\partial t$. In the general case when there is viscosity and heat transfer, the right-hand side can be written, in the non-relativistic limit, as 
\begin{equation}
  {T^{0\nu}_{\rm m}}_{,\nu} = \frac{\partial}{\partial t}(\frac{1}{2}\rho v^2 + u) + \mathbf\nabla . [\rho\mathbf v (\frac{1}{2} v^2 + w) - \mathbf v . \mathbf\sigma ' + \mathbf J],
\end{equation}
where $w$ is the specific enthalpy, $u$ is the internal energy density, $\mathbf J$ is the heat flux, which can generally be written as $-\kappa\mathbf\nabla T$, $T$ is the temperature and $\kappa$ is the thermal conductivity. Finally, $(\mathbf v . \mathbf\sigma')_k$ stands for $v_i \sigma ' _{ik}$, where $\sigma '$ is the viscous stress tensor, \citep{Landaufluids}. As we stated above, we neglect the velocity of the fluid, so obtain
\begin{equation}
  {T^{0\nu}_{\rm m}}_{,\nu} = \frac{\partial u}{\partial t} + \mathbf{\nabla\cdot J}.
\end{equation}

A note of caution regarding the internal energy is in order. We understand, as usual, ``internal energy'' as the energy that can be exchanged by the system in the processes considered (heat exchange, radiative transfer, etc.), which will differ from what we understand by ``rest mass'', which is the ``non-convertible energy''. If the scalar field can change the effective electric charge, then it can alter the electromagnetic contribution to the rest mass, and consequently, this contribution will be no longer ``rest mass'', but ``internal energy''. 
  
   The time variation in the internal energy $u$ will have two contributions: one corresponding to the cooling process $\frac{\partial u}{\partial t}|_{cooling}$ and another one related to the interaction with the scalar field  $\frac{\partial\sigma_\mu}{\partial t}$. This last term accounts for the dependence of the bulk of matter on the scalar field, which is mainly given by the electromagnetic contribution to the nuclear mass. Equation (\ref{div:Tm0}) then will finally read

\begin{equation}
  \begin{split}
  \frac{\partial u}{\partial t}|_{cooling}+ \frac{\partial \sigma_{\mu}}{\partial t}+ \mathbf{\nabla\cdot J} =& -\frac{\bm{B}^2}{4\pi} \dot{\psi}e^{-2\psi}  
- \frac{e^{-2\psi}\mathbf B^2\dot\psi}{4\pi}\\ 
  &- e^{-2\psi}\mathbf\nabla\psi . \mathbf S
-\dot\psi\frac{\partial\sigma}{\partial\psi}. 
  \end{split}
\end{equation}

Since the scalar field is space independent, and given that the electromagnetic energy of matter is mainly accounted for the nuclear content, we assume that the following condition $\frac{\partial \sigma}{\partial\psi}-\frac{\partial\sigma_{\mu}}{\partial\psi}\approx 0$ is fulfilled. We consequently obtain

\begin{equation}
  \frac{\partial u}{\partial t}|_{cooling}+ \mathbf{\nabla\cdot J} \approx - \frac{e^{-2\psi}\mathbf B^2\dot\psi}{2\pi} \label{cooling}. 
\end{equation}

This equation becomes clearer if we make a trivial change to produce
\begin{equation}
  \mathbf{\nabla\cdot J} = -\frac{e^{-2\psi}\mathbf B^2\dot\psi}{2\pi} - \frac{\partial u}{\partial t}|_{cooling} \label{cooling2},
\end{equation}
which clearly shows that besides the standard cooling mechanism of the body, there is a contribution from the partial release of the magnetic energy injected by the scalar field. We define

\begin{equation}
\varepsilon_{a}= 2\frac{e^{-2\psi}\mathbf B^2\dot\psi}{M_{a}4\pi}\approx  2\frac{\dot\alpha}{\alpha}\frac{\mathbf B^2}{8\pi M_{a}},\label{equ:TV-E:alfa:num}
\end{equation}
to be equal to twice the energy production per mass unit of any material substance $a$ (using the approximation, $e^{-2\psi}\to 1$ when $\psi<<1$).

 We now consider our main physical assumption: {\it the cooling term is not modified by the scalar field}. The reasons for this assumption are fold: 1) as we have just shown, the electrostatic energy ``injected'' by the scalar field remains within the bulk matter (the cancellation of terms occurring as seen in Eq.(\ref{cooling})), and 2) the thermal evolution should not change given the high thermal conductivity of the Earth and white dwarfs considered in this work. Thus, we expect the magnetic energy excess to be radiated away, increasing the heat flux $\mathbf J$ as shown in Eq.(\ref{cooling2}).

\section{The electromagnetic energy of matter}
\label{sec:Eem-Mat}

As we mentioned in the Sect. \ref{sec:Edot}, the only ``input'' is that derived from the magnetic field. Stationary electric currents generated by charged particles and their static magnetic moments, and quantum fluctuations of the number density are responsible for the generation of magnetic fields in quantum mechanics. These contributions have been studied and calculated by \citet{Haugan:1977px} and \citet{WillTh} from a minimal nuclear shell model using the following analysis (for more details see \citet{KV1-09}).

 The total magnetic energy of the nucleus can be written as

\begin{equation}
  \label{eq:Mag:Ener:Expr}
  E_m \simeq\frac{1}{2c^2}\sum_\alpha \int d\bm{x} d\bm{x}' \frac{\left\langle0\right\vert
    \bm{j}(\bm{x})\left\vert \alpha\right\rangle \bm{\cdot}
    \left\langle\alpha\right\vert \bm{j}(\bm{x'}) \left\vert0\right\rangle }%
  {\vert \bm{x} - \bm{x}' \vert},
\end{equation}
where $\alpha$ runs over a complete set of eigenstates of the nuclear
Hamiltonian $H$. Neglecting the momentum dependence of the nuclear potential and assuming a constant density within the nucleus, we obtain the result
\begin{equation}
  \label{eq:Bek:Curr:Prod}
  \left\langle0\right\vert \bm{j}(\bm{x}) \left\vert\alpha\right\rangle \bm{\cdot}
  \left\langle\alpha\right\vert \bm{j}(\bm{x'}) \left\vert0\right\rangle \simeq
  \frac{\left\vert d_{0\alpha} \right\vert^2}{\hbar^2}
  \frac{E_{0\alpha}^2}{V_N^2} \cos\theta, 
\end{equation}
where $\bm{d}_{0\alpha}$ is the dipole density, $V_N = \frac{4\pi}{3} R_N^3 $ is the nuclear volume, and  $\theta$ is the angle between $\hat{\bm{x}}$ and
$\hat{\bm{x}}'$. Hence,

\begin{equation}
  \label{eq:Mag:Ener:Expr:2}
  E_m \simeq \frac{\sum_a E_{0\alpha}^2 \left\vert d_{0\alpha}\right\vert^2}{2\hbar^2c^2}
        \frac{\int d\bm{x}d\bm{x}' \frac{\cos\theta}{\vert \bm{x} -
            \bm{x}' \vert}}{V_N^2}.
\end{equation}
 In the last equation, $\frac{\int d\bm{x}d\bm{x}' \frac{\cos\theta}{\vert \bm{x}-\bm{x}' \vert}}{V_N^2}$ is equal to $\frac{3}{5R_N}$, and the first term can be computed from the connection between the strength function and the
photoabsorption cross-section

\begin{equation}
  \label{eq:Def:Strength:Function}
  \sigma_{0\alpha} = \frac{4\pi}{\hbar c} E_{\alpha0} \vert d_{\alpha0} \vert^2.
\end{equation}

From this, we easily find that
\begin{equation}
  \begin{split}\label{eq:Mom:Strength}
    \sum_a E_{\alpha0}^2 \vert d_{\alpha0} \vert^2 =& \frac{\hbar c}{4\pi}
  \frac{\int E\sigma(E)dE}{\int\sigma(E)dE} \cdot  \int\sigma(E)dE\\
    &= \bar{E} \int\sigma(E)dE,
  \end{split}
\end{equation}
where $\bar{E} \sim \unit[25]{MeV}$ is the mean absorption energy,
which is roughly independent of $A$ (number of nucleons). 

The cross-section satisfies the Thomas-Reiche-Kuhn 
sum rule
\begin{equation}
  \label{eq:TRK:Sum:Rule}
  \int\sigma(E)dE = (1+x)\frac{2\pi^2e^2\hbar}{mc} \frac{NZ}{A} \simeq
  (1+x)\unit[15]{MeV\,mbarn} A,
\end{equation}
where $x \sim 0.2$ takes into account exchange and velociy dependence of
nuclear interactions. Combining Eqs.\eqref{eq:Mag:Ener:Expr:2},
\eqref{eq:Mom:Strength}, and \eqref{eq:TRK:Sum:Rule}, we obtain

\begin{equation}
  \begin{split}\label{eq:Bek:Mag:Ener}
    E_m =& \int d^3x \frac{B^2}{8\pi}  \simeq \frac{1}{2c^{2}} \int d^3x d^3x'
  \frac{\bm{j}(\bm{x})\cdot\bm{j}(\bm{x}')}{\mid \bm{x} - \bm{x}'\mid }\\ 
    &\simeq \frac{3}{20\pi} \frac{\bar{E}}{R(A)\hbar c} \int\sigma dE,
  \end{split}
\end{equation}
where $R(A)$ is the nuclear radius. These quantities have the following approximate
representation
\begin{align}
  \label{eq:Bek:Mag:E:Par}
  R(A) &=  \unit[1.2 A^{\frac{1}{3}}]{fm}, &  \int\sigma dE &\simeq  \unit[1.6 A]{MeV\; fm^2}.
\end{align}
The fractional contribution of the magnetic energy to rest mass energy is then
\begin{equation}
\zeta(A) \simeq \frac{E_{m_A}}{m_{A}c^2}\approx 8.60465\times 10^{-6}A^{-1/3}.
\end{equation}

Table \ref{tab:val:zeta} shows typical values of $\zeta(A)$ computed using the semi-empirical mass formula and the contribution of neutrons and protons.
\begin{table}[H]
 \begin{center}
  \begin{tabular}{|*{2}{>{$}c<{$}|}}
	\hline
 	\mbox{Nucleus} & 10^6\zeta\\
 	\hline
 	{\rm He}^4_2 & 5.42\\
 	{\rm C}^{12}_6 & 3.76\\
 	{\rm O}^{16}_8 & 3.41\\
 	{\rm Si}^{28}_{14} & 2.83\\
 	{\rm Fe}^{56}_{26} & 2.25\\
 	\hline
  \end{tabular}
 \end{center}
 \caption{$\zeta$ values for typical indoor stellar and planetary elements}\label{tab:val:zeta}
\end{table}

\section{The Earth heat flux}
\label{sec:CoolEarth}
 
  There are several models that attempt to explain the average rate of secular cooling of the Earth in terms of variations in the composition of the mantle melts through time \citep{Lab07}. Constraints on these theories are set by terrestrial heat flow measurements on the surface.

 The contribution of $\dot{\alpha}/\alpha$ to the heat flux can be calculated using the global heat balance for the Earth, assuming that the \emph{machian} contribution $H_{C}$ is the only extra energy production,
\begin{equation}
  \label{eq:EngEarthbalance}
M_{E}C_{p}\frac{dT_{m}}{dt}=-Q_{tot}+H_{C}+H_{G},
\end{equation}
where $M_{E}$ is the Earth's mass, $C_{p}$ is the average heat capacity of the planet, $T_{m}$ is the mantle potential temperature, and $H_{G}$ represents the heat generated by radioactive isotopes. The total heat loss $Q_{tot}$ can be written as the sum of two terms, one that comes from the loss of heat in the oceans $Q_{oc}$, and the other from continental heat loss $Q_{cont}$. Using the results obtained by \citet{Lab07}, we rewrite the total heat loss as $Q_{tot}\approx MC_{p}\lambda_{G}T_{m}$, where $\lambda_{G}$ is the timescale constant for the secular Earth's cooling. Assuming that the most abundant elements of the Earth are oxygen, silica, and iron, the ``extra'' energy contribution can be written as
\begin{equation}
H_{C}=\bar{\zeta} c^2H_{0}\frac{\dot{\alpha}}{\alpha H_{0}},
\end{equation}
where $\bar{\zeta}$ is the mass-weighted averaged of the parameter $\zeta(A)$.
\begin{table}[H]
 \begin{center}
  \begin{tabular}{|*{2}{>{$}c<{$}|}}
	\hline
 	\mbox{Parameter} & Value\\
 	\hline
        {\rm H_{0}} & \unit[2.5\times 10^{-18}] {s^{-1}}\\
 	{\rm M_{E}} & \unit[5.94\times 10^{24}] {Kg}\\
 	{\rm C_{P}} & \unit[1200] {J/Kg-K}\\
 	{\rm \bar{\zeta}} & 2.75\times 10^{-6}\\
        {\rm \lambda_{G}} &  \unit[0.1] {Gyr^{-1}}\\
 	\hline
  \end{tabular}
 \end{center}
 \caption{Values of parameters}\label{tab:val:par}
\end{table}
  From Eq.\eqref{equ:Int1ra}, we can describe the extra contribution as a function of time, writing $\frac{a(t)}{a_{0}}$ as a power series, \citep{Weinberg}
\begin{equation}
\frac{a(t)}{a_{0}}\approx 1+H_{0}dt-\frac{q_{0}}{2}(H_{0}dt)^2+\frac{j_{0}}{6}(H_{0}dt)^3+\cdots\label{expanda}
\end{equation}
 and then making a Taylor series expansion up to third order of $H_{C}$. Replacing this \emph{machian} contribution in Eq.\eqref{eq:EngEarthbalance}, solving it, and using the parameter's values from Table \ref{tab:val:par}, we find that the cosmological perturbation of the mantle's temperature $\Delta T_{m}$ in terms of the time interval $\Delta t$ and $\frac{\dot{\alpha}}{\alpha H_{0}}$ is given by

\begin{equation}
  \begin{split}
    &\Delta T_{m}(t) =\unitfrac[2.43\times 10^5]{K}{Gyr} \frac{\dot{\alpha}}{H_0\alpha}(\Delta t)^3\\
    &-\unitfrac[3.78\times10^6]{K}{Gyr} \frac{\dot{\alpha}}{H_0\alpha}(\Delta t)^2\\
    &+\unitfrac[3.05\times 10^7]{K}{Gyr} \frac{\dot{\alpha}}{H_0\alpha}\Delta t.
  \end{split}\label{equ:temp1}
\end{equation}

 According to \citet{Lab07}, \emph{the total amount of cooling experienced by the Earth after an initial magma ocean phase cannot exceed $\unit[200] {K}$}. Hence, in the past $\unit[2.5] {Gyr}$, $\Delta T_{m}<\unit[200] {K}$. With these restrictions, we obtain a bound for the time variation in $\alpha$ of,  
\begin{equation}
\left\vert\frac{\dot{\alpha}}{H_0\alpha} \right\vert_0 <1.93\times10^{-6}.\label{resultsE}
\end{equation}
 Using this result in Eq.(\ref{equ:Pred}), we find that,
\begin{equation}
\left(\frac{\ell_B}{\ell_P}\right)^2<0.15,\qquad
\frac{\ell_B}{\ell_P}<0.39.\label{equ:Pred:tierra}
\end{equation}

  A different bound can be obtained by observing that the total radiated power of the Earth $Q_{tot}$ can be explained by radiactive decay to within twenty per cent (\citet{Lab07}). The most recent data was estimated from an adjustment made with 38347 measurements. The methodology was to use a half-space cooling approximation for hydrothermal circulation in young oceanic crust; and in the remainder of the Earth's surface, the average heat flow of various geological domains was estimated as defined by global digital maps of geology, and then a global estimate was made by multiplying the total global area of the geological domain, \citep{Davis10}. 

 The result shows that $Q_{tot}\approx \unit[47] {TW}$ ( in Table \ref{tab:val:hf} this estimate is separated into continental and oceanic contributions). Therefore,
\begin{equation}
\left\vert Q_{mach}\right\vert=\left\vert M_{E}C_{P}\lambda_{G}T_{m}(t)\right\vert <0.2Q_{tot}.\label{Q:mach}
\end{equation}

\begin{table*}[!ht]
\renewcommand{\arraystretch}{1.3}
\begin{center}
\begin{tabular}{|c|c|c|c|}
\hline
 Part of the Earth  & Area ($\unit[10^{14}]{m^2}$)   & Heat flow ($\unit{TW}$) &  Mean heat flow ($\unit{\frac{mW}{m^2}}$) \\ \hline
Continent & 2.073 & 14.7  &  70.9 \\\hline
 Ocean & 3.028& 31.9 & 105.4\\\hline
 Global Total & 5.101 & 46.7 & 91.6 \\\hline
\end{tabular}
\end{center}
\caption{Summary of the heat flow from \citet{Davis10} preferred estimates}
\label{tab:val:hf}
\end{table*}

In an interval of $\unit[2.5] {Gyr}$, then we find that 
\begin{equation}
\left\vert\frac{\dot{\alpha}}{H_0\alpha} \right\vert_0 <3.98\times10^{-6},\label{resultsE2}
\end{equation}  
and
\begin{equation}
\left(\frac{\ell_B}{\ell_P}\right)^2<0.31,\qquad
\frac{\ell_B}{\ell_P}<0.55.\label{equ:Pred:tierraII}
\end{equation}

\section{The heat fluxes of the outer planets}
\label{sec:outerplanets}

Jupiter, Saturn, Uranus, and Neptune are often called gas giants. They are massive planets with a thick atmosphere and a solid core. Jupiter and Saturn are composed primarily of hydrogen and helium, while Uranus and Neptune are sometimes called ice giants, as they are mostly composed of water, ammonia, and methane ices.
By comparising the observed bolometric temperatures of giant planets with those expected when the planets are in thermal equilibrium with incident solar radiation, it is clear that all of these planets except for Uranus have a significant internal heat source, \citep{Irwin01}.

In the case of Jupiter, the residual primordial heat emitted is caused by the continued cooling and shrinking of the planet via the \emph{Kelvin-Helmholtz mechanism}. 

Saturn must have also started out hot inside like Jupiter as the result of its similar formation. But being somewhat smaller and less massive, Saturn was not as hot in the beginning of its life and has had time to cool. As a result, this planet has lost most of its primordial heat and there must be another source of most of its internal heat. This excess heat is generated by the precipitation of helium into its metallic hydrogen core. The heavier helium separates from the lighter hydrogen and drops toward the center. Small helium droplets form where it is cool enough, precipitate or rain down, and then dissolve at hotter deeper levels. As the helium at a higher level drizzles down through the surrounding hydrogen, the helium converts some of its energy into heat, \citep{Lang09}.  

The low value of Uranus' internal heat is still poorly undertood. One suggestion is that chemical composition gradients may act as inhibitors of heat transport from the planet's hot interior to the surface. Another hypothesis is that it was hit by a supermassive impactor that caused it to expel most of its primordial heat, leaving it with a depleted core temperature. Uranus has as much as $4M_{\oplus}$ of rocky materials hence, part of the internal flux ($\approx 0.02Wm^{-2}$) comes from radioactive decay; \emph{Kelvin-Helmholtz mechanism} would also be expected.

Although Neptune is much farther from the Sun than Uranus, its thermal emission is almost equivalent. Several possible explanations have been suggested, including radiogenic heating from the planet's core, conversion of methane under high pressure into hydrogen, diamond and longer hydrocarbons (the hydrogen and diamond would then rise and sink, respectively, releasing gravitational potential energy), and convection in the lower atmosphere that causes gravity waves to break above the tropopause \citep{Fortney-Hubbard04,Hubbard78}.

  We calculate the heat fluxes $ J_{\zeta_{i}}$ for each planet using the equation of heat conduction
\begin{equation}
  \frac{1}{r^2} \frac{d\ }{dr} \left(Kr^2\frac{dT}{dr}\right) =
  -\varepsilon\rho, \label{equ:CondCal}
\end{equation}
where $K$ is the effective thermal conductivity of the planet material. The heat flux is
\begin{equation}
 \mathbf J= - K\frac{dT}{dr}.    \label{def:j}
\end{equation}

If $\bar{\varepsilon}$ is the planet mean heat production, which is estimated from the results in Table \ref{tab:val:zeta} according to the chemical composition of each planet, then
\begin{equation}
  J(r) =  - K\frac{dT}{dr} = \frac{1}{r^2}
  \int_0^\infty \varepsilon(r')\rho(r') dr' =
  \bar{\varepsilon}\frac{m(r)}{4\pi r^2}, \label{equ:j(r)}
\end{equation}
hence the surface flux is
\begin{equation}
  \mathbf J_{\zeta_{i}} = -\left. K\frac{dT}{dr}\right\vert_{\zeta_{i}} =
  \bar{\varepsilon}\frac{m(R_i)}{4\pi R_i^2}, \label{equ:j_R}
\end{equation}
which is the fundamental equation. Thus, we compare the results of $ J_{\zeta_{i}}$ with the observed fluxes obtained with Voyager 1, 2, and Cassini, \citep{Pearl90}.

\begin{table*}[!ht]
\renewcommand{\arraystretch}{1.3}
\begin{center}
\begin{tabular}{|c|c|c|c|c|}
\hline
 Planet  & $J_{obs}$ ($\unit{W/m^2}$)   & M ($\unit{Kg}$) & R ($\unit{m}$) & $J_{\zeta_{i}}$ ($\unit{W/m^2}$) $\frac{\dot{\alpha}}{H_0\alpha}$ \\ \hline
Jupiter & $5.44\pm0.43$ & $1.90\times10^{27}$ & $7.14\times10^{7}$ & $6.35\times10^{4}$\\\hline
 Saturn & $2.01\pm0.14$ & $5.68\times10^{26}$ & $6.03\times10^{7}$ & $2.71\times10^{4}$\\\hline
 Uranus & $0.042\pm0.047$ & $8.68\times10^{25}$ & $2.556\times10^{7}$ & $2.08\times10^{4}$\\\hline
Neptune & $0.43\pm0.09$ & $1.02\times10^{26}$ & $2.47\times10^{7}$ & $3.44\times10^{4}$ \\\hline
\end{tabular}
\end{center}
\caption{The observed heat flux, mass, radius, and calculated heat flux of the outer planets}
\label{tab:OuterHeatFlow}
\end{table*}

 The ``$3\sigma$'' upper bounds and the corresponding ``$\left(\ell_B/\ell_P\right)$'' bounds are

\begin{table}[H]
  \begin{center}
    \begin{tabular}{|l|*{2}{>{$}c<{$}|}}
	\hline
	Planet & \left\vert \frac{\dot{\alpha}}{H_0\alpha} \right\vert  & \left(\ell_B/\ell_P\right)
	\\
	\hline
	Jupiter & 2.04\times10^{-5}  & 1.25\\
	Saturn & 1.55\times10^{-5} & 1.09\\
	Uranus & 6.75\times10^{-6} & 0.72\\
	Neptune & 7.85\times10^{-6} & 0.78\\
        Earth$^{(1)}$ & 1.93\times10^{-6} & 0.39\\
        Earth$^{(2)}$ & 3.98\times10^{-6} & 0.55\\
	\hline
  \end{tabular}
  \end{center}
  \caption{Bounds from the outer planets and the Earth ($^{(1)}$ results from Eqs.(\ref{resultsE}) and (\ref{equ:Pred:tierra}) and $^{(2)}$ from Eqs.(\ref{resultsE2}) and (\ref{equ:Pred:tierraII})).}
  \label{tab:OuterHeatFlowresults}
\end{table}

\section{Conclusions}
\label{sec:conclusions}
  The energy exchange with ordinary matter in alternative theories with new fields such as Beckenstein's theory is a quite undeveloped field of subject. Using the field equations and general hypothesis of the theory we have derived the energy transfer between matter and fields. The Hypothesis \ref{item:Einstein} in Sect. \ref{sec:alfadot} is a probable key, because it states that the matter energy momentum tensor is the quantity that has to be added to the field sector in order to make the total tensor divergence free. We have also assumed that dark matter is electrically neutral, have neglected the motion of matter in the bodies considered, and have found that the dynamical feature of the electric charge makes the atomic electromagnetic energy part of the internal energy of the system. Eq.(\ref{cooling}) shows that there is another contribution to the heat current in addition to the cooling of matter, which is given by the time variation in the scalar field and the magnetic content of matter. We have also justified our assumption that the matter cooling rate is not modified by the scalar field. Finally, using a minimal nuclear shell model we estimated the magnetic energy content of matter, thus permitting us to quantify the anomalous heat flux in terms of the fundamental parameters of the theory and the chemical composition of the body.  

  Our tightest constraint was obtained by analyzing the geothermal aspects of the Earth, which are naturally the most clearly understood and reliably measured of our solar system, and the surface heat flux is very low. Our bounds are comparable with that obtained in the laboratory by combining measurements of the frequences of \rm{Sr} (\citet{Blatt08}), \rm{Hg+} (\citet{Fortier06}), \rm{Yb+} (\citet{Peik04}) and \rm{H} (\citet{Fisher04}) relative to Caesium \citep{Barrow10,Uzan10}, and only one order of magnitude weaker than Oklo's, which is the most stringent constraint on $\alpha$'s time variation up to date \citep{Uzan10,Fujii2000}), and another found from measurements of the ratio of \rm{Al+} to \rm{Hg+} optical clock frequencies over a period of a year \citep{Rosen08,Barrow10}(see Table \ref{tab:sum:alfa}). The constraints we found depend on the cooling model of the Earth, but there is a general agreement about the mechanisms behind it (\citet{jessop90}). The outer planets have provided us with  additional constraints, which are between the same and one order of magnitude weaker than Earth's, but are nevertheless valuable, as the chemical composition and cooling mechanisms differ widely from our planet. The data set considered here is able to place independent constraints on the theory parameters. This analysis may be applied to other theories with extra fields that introduce extra ``internal energies'' to matter. We will report on further work in future publications. 

\begin{table}[H]
 \begin{center}
  \begin{tabular}{|*{3}{>{$}c<{$}|}}
	\hline
 	\mbox{Constraint} & \left\vert \frac{\dot{\alpha}}{\alpha} \right\vert (\unit{yr^{-1}}) & Reference\\
 	\hline
        {\rm Clocks} \rm{Cs} & (3.3\pm 3.0)\times10^{-16} & (1)\\
        {\rm Clocks} \rm{Hg} & (5.3\pm 7.9)\times10^{-17} & (1)\\
 	{\rm Oklo} & (2.50\pm 0.83)\times10^{-17} & (2)\\
 	{\rm J_{\oplus}} & 1.52\times10^{-16}  & (3)\\
	{\rm J_{\oplus}}II   & 3.14\times10^{-16} & (4)\\
	{\rm J_{Jup}} & 1.61\times10^{-15} & (5)\\
	{\rm J_{Sat}} & 1.22\times10^{-15} & (5)\\
	{\rm J_{Ur}} & 5.32\times10^{-16} & (5)\\
	{\rm J_{Nep}} & 6.19\times10^{-16} & (5)\\

 	\hline
  \end{tabular}
 \end{center}
 \caption{The table comparises different kinds of constraints, the value of
  $\frac{\dot \alpha}{\alpha_0}$, and the reference. References (1) \citet{Barrow10}; (2)\citet{Fujii2000}; (3) Eq.\ref{resultsE}; (4) Eq.\ref{resultsE2}; (5) Table\ref{tab:OuterHeatFlowresults}}\label{tab:sum:alfa}
\end{table}

\bibliography{Qdot.bib} 
\bibliographystyle{aa}

\end{document}